\documentclass[aps,prl,nobibnotes,superscriptaddress,twocolumn,10pt]{revtex4-1}

\usepackage{graphicx} 	
\usepackage{dcolumn} 	
\usepackage{bm}   		
\usepackage{amssymb}   	
\usepackage[separate-uncertainty=true]{siunitx}		
\usepackage[svgnames]{xcolor}	
\usepackage[version=3]{mhchem}	

\usepackage{titlesec}
\titlespacing*\section{0pt}{2\parindent}{.5\parindent}
\titlespacing*\subsection{0pt}{\parindent}{0pt}
\titlespacing\subsubsection{0pt}{0pt}{5pt}[]
\titleformat{\section}[display]{\bfseries}{\thesection}{0pt}{\uppercase}
\titleformat{\subsection}[display]{\bfseries}{\thesubsection}{0pt}{}
\titleformat{\subsubsection}[runin]{\normalfont\bfseries}{\thesubsubsection}{}{}[:]

\usepackage{xr} 
\usepackage[hidelinks]{hyperref}
\usepackage[caption=false]{subfig}
\usepackage[nameinlink,capitalise]{cleveref}
\usepackage{lipsum}

\newcommand{\po}{$p_{\ce{O2}}$}
\DeclareSIUnit\unitcell{uc}

\graphicspath{{figs/}}

\graphicspath{{figs/}}

\begin{document}

\title{Charge doping and large lattice expansion in oxygen-deficient heteroepitaxial \ce{WO3}}

\author{Giordano Mattoni\thanks{Contact Address: g.mattoni@tudelft.nl}}
\affiliation{Kavli Institute of Nanoscience, Delft University of Technology, 2628 CJ Delft, Netherlands}

\author{Alessio Filippetti}
\affiliation{Dipartimento di Fisica, Universit\`{a} di Cagliari, and CNR-IOM, Istituto Officina dei Materiali, Cittadella Universitaria, Cagliari, Monserrato 09042-I, Italy}

\author{Nicola Manca}
\affiliation{Kavli Institute of Nanoscience, Delft University of Technology, 2628 CJ Delft, Netherlands}

\author{Pavlo Zubko}
\affiliation{London Centre for Nanotechnology, University College London, 17--19 Gordon Street, London WC1H 0HA, UK}

\author{Andrea D. Caviglia}
\affiliation{Kavli Institute of Nanoscience, Delft University of Technology, 2628 CJ Delft, Netherlands}

\begin{abstract}
Tungsten trioxide is a versatile material with widespread applications ranging from electrochromic and optoelectronic devices to water splitting and catalysis of chemical reactions.
For technological applications, thin films of \ce{WO3} are particularly appealing, taking advantage from high surface-to-volume ratio and tunable physical properties.
However, the growth of stoichiometric, crystalline thin films is challenging because the deposition conditions are very sensitive to the formation of oxygen vacancies.
%
In this work, we show how background oxygen pressure during pulsed laser deposition can be used to tune the structural and electronic properties of \ce{WO3} thin films.
By performing X-ray diffraction and low-temperature transport measurements, we find changes in \ce{WO3} lattice volume up to \SI{10}{\percent}, concomitantly with an insulator-to-metal transition as a function of increased level of electron doping.
We use advanced ab initio calculations to describe in detail the properties of the oxygen vacancy defect states, and their evolution in terms of excess charge concentration.
Our results depict an intriguing scenario where structural, electronic, optical, and transport properties of \ce{WO3} single-crystal thin films can all be purposely tuned by a suited control of oxygen vacancies formation during growth.

\end{abstract}

\date{\today}
\maketitle

\section{Introduction}

The tungsten oxide \ce{WO3} holds a special place in the family of complex oxides, since its perovskite \ce{ABO3} crystal structure has an empty \ce{A}-site.
This characteristic determines an open crystalline structure, which is prone to host interstitial species which act as dopants for the otherwise insulating material \cite{
	granqvist2000electrochromic,
	hirai2011superconductivity,
	haldolaarachchige2014superconducting,
	soma2016epitaxial,
	hamdi2016first
}.
For these reasons \ce{WO3} finds wide use in electrochromic, optoelectronic and gas sensing applications \cite{
	deb2008opportunities,
	long2015synthesis,
	cong2016tungsten
}.
Most works so far focused on thick films, amorphous layers and nanorods \cite{
	legore1997controlled,
	tagtstrom1999chemical
}.
Only recently the growth of crystalline thin films has been demonstrated by means of several techniques such as sputtering, molecular beam epitaxy and pulsed laser deposition \cite{
	du2014strain,
	leng2015epitaxial,
	kalhori2016morphology,
	herklotz2017symmetry
}.
Because \ce{WO3} structure and electronic properties are very sensitive to oxygen stoichiometry \cite{
	sahle1983electrical,
	he2013memristive,
	meng2015electrolyte,
	altendorf2016facet
}, a precise control of oxygen partial pressure during the growth process is crucial to obtain high quality thin films.

In this work, we study the effects of oxygen pressure during pulsed laser deposition of \ce{WO3} thin films, and show how it modifies the material structural and electronic properties.
X-ray diffraction measurements reveal that heteroepitaxial \ce{WO3} thin films are in a tetragonal phase, where the out-of-plane lattice parameter can be gradually tuned up to \SI{10}{\percent}, changing from \SI{3.7}{\angstrom} measured in the most stoichiometric compound to \SI{4.1}{\angstrom} in the most oxygen deficient case.
By performing transport measurement, we find a semiconducting trend characterised by an activated transport regime, with an energy gap that vanishes for increased level of oxygen vacancies.
These findings are corroborated by \textit{ab initio} calculations, showing that oxygen vacancies form in-gap states, effectively donating electron carriers and increasing \ce{WO3} lattice volume.
Our results show how to obtain \ce{WO3} thin films with high crystal quality and controlled electronic properties.

\begin{figure}[tb]
\includegraphics[page=1,width=1\linewidth]{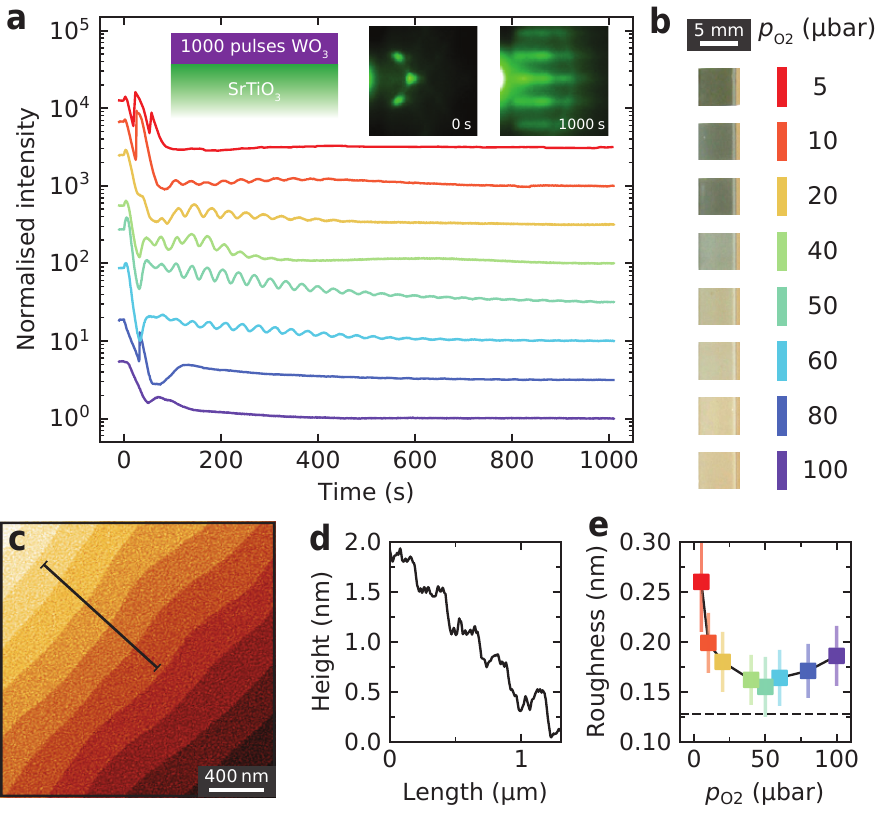}

\subfloat{\label{fig:RHEED}}
\subfloat{\label{fig:Microscope}}
\subfloat{\label{fig:WO3_AFM}}
\subfloat{\label{fig:AFM_profile}}
\subfloat{\label{fig:Roughness}}

\caption{\textbf{Growth of \ce{WO3} thin films.}
\protect\subref{fig:RHEED} RHEED intensity oscillations during film growth at different \po and
(inset) heterostructure schematic, initial and final RHEED diffraction patterns at \po$=\SI{50}{\micro\bar}$.
\protect\subref{fig:Microscope} Photograph of the samples,
\protect\subref{fig:WO3_AFM} surface topography by atomic force microscopy with
\protect\subref{fig:AFM_profile} line profile along the black line and
\protect\subref{fig:Roughness} surface roughness calculated as the root mean square of the topographic signal.
}

\label{fig:Growth}

\end{figure}

\begin{figure*}[t]
\includegraphics[page=2,width=.7\linewidth]{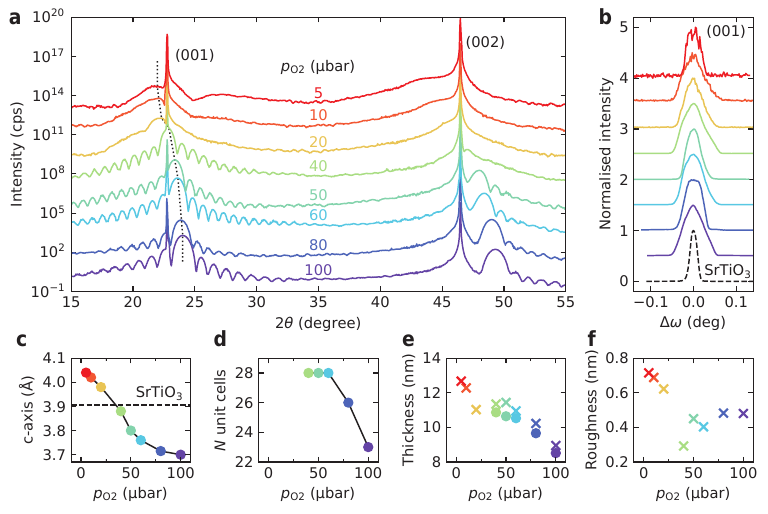}

\subfloat{\label{fig:WO3_T2T}}
\subfloat{\label{fig:WO3_Rocking}}
\subfloat{\label{fig:caxis}}
\subfloat{\label{fig:Nuc}}
\subfloat{\label{fig:XRD_thickness}}
\subfloat{\label{fig:XRD_roughness}}

\caption{\textbf{X-ray diffraction characterisation.}
\protect\subref{fig:WO3_T2T} $\theta$--$2\theta$ scans around the (001) and (002) peaks of the \ce{SrTiO3} substrate.
The dotted line highlights the position of the \ce{WO3} (001) peak.
\protect\subref{fig:WO3_Rocking} Rocking curves around \ce{WO3} (001) peak compared to the substrate (001) peak (black dashed line).
\protect\subref{fig:caxis} Film c-axis parameter and substrate lattice constant (dashed line),
\protect\subref{fig:Nuc} number of \ce{WO3} unit cells extracted by simulating the finite size oscillations with a kinematic scattering model.
\protect\subref{fig:XRD_thickness} Film thickness and
\protect\subref{fig:XRD_roughness} interface roughness from reflectivity measurements (crosses) and the $\theta$--$2\theta$ data (circles).
}

\label{fig:WO3_XRD}

\end{figure*}

\section{Film growth}
To study the effect of background oxygen pressure (\po) on \ce{WO3} thin films grown by pulsed laser deposition (PLD), we prepare a series of samples deposited on top of \ce{TiO2}-terminated \ce{SrTiO3} (001) substrates.
We use a laser fluence of \SI{1}{\joule\per\square\centi\metre}, a repetition rate of \SI{1}{\hertz}, a target-to-sample distance of \SI{55}{\milli\metre} and a fixed deposition duration of 1000 laser pulses.
The growth temperature is \SI{500}{\celsius}, while \po is varied in the range \SIrange{5}{100}{\micro\bar}.
The plasma plume appears more diffused at lower \po and more intense at higher \po (\cref{fig:Plumes}).
This is because the oxygen pressure in the PLD chamber strongly influences the plume dynamics, enhancing scattering of the ablated species and thermalisation with the background gas.
As a consequence, their oxidation state and kinetic energy can be modified, so that the stoichiometry of the deposited material strongly depends upon \po \cite{
	wicklein2012pulsed,
	groenendijk2016epitaxial
}.
We monitor the growth \textit{in situ} by reflection high-energy electron diffraction (RHEED) and observe clear intensity oscillations (\cref{fig:RHEED}) when \po is between \SIrange{10}{60}{\micro\bar}.
The RHEED pattern evolves from three well defined diffraction points, typical for \ce{SrTiO3} single crystals \cite{
	haeni2000rheed
}, to a series of stripes that indicate bidimensional film growth.
As shown in the photograph of \cref{fig:Microscope}, the sample colour is also affected by the oxygen pressure, and it gradually changes from transparent to dark grey with lowering \po.
All the deposited \ce{WO3} films have a step-and-terrace surface morphology with single unit cell steps (\cref{fig:WO3_AFM,fig:AFM_profile}).
This structure mimics the underlying \ce{SrTiO3} substrate, indicating uniform growth (\cref{fig:AFM_many}).
The samples have very low surface roughness (\cref{fig:Roughness}) ranging from \SIrange{0.15}{0.25}{\nano\metre}, which is comparable to what we measure in a pristine substrate (\SI{0.13}{\nano\metre}).
Interestingly, the roughness is minimal at \po$=\SI{50}{\micro\bar}$, pressure for which the RHEED oscillations during growth are more accentuated.
This characterisation indicates that for all \po in the explored range the films grow with a smooth surface morphology.

\section{Experimental characterisation}
\subsection{X-ray diffraction}
To evaluate the crystal quality of our \ce{WO3} thin films, we perform X-ray diffraction measurements.
\Cref{fig:WO3_T2T} shows $\theta$--$2\theta$ scans around the sharp (001) and (002) peaks of the \ce{SrTiO3} substrate.
The \ce{WO3} film grown at \po$=\SI{100}{\micro\bar}$ presents peaks at $2\theta=\ang{24}$ and \ang{49}, surrounded by neat finite size oscillations that demonstrate a high crystalline quality.
No additional diffraction peaks are observed, indicating that the thin films are in a single crystal phase.
For lower \po, the peaks and finite size oscillations gradually shift to lower $2\theta$ angles.
At \po$\leq\SI{20}{\micro\bar}$ the finite size oscillations become less defined, concurrently with a broadening of the diffraction peaks.
Such signal degradation is usually determined by a decreased crystal quality, similarly to what has previously been observed for highly doped \ce{WO3}
\cite{
	he2016atomistic,
	wang2016electron
}.
We evaluate the presence of defects in \ce{WO3} by measuring rocking curves around the film (001) peak.
In \cref{fig:WO3_Rocking} we find for all samples a sharp peak with full width at half maximum between \ang{0.02} and \ang{0.03}, very close to the value found for the underlying \ce{SrTiO3} substrate (\ang{0.01}).
The sharp rocking curves indicate that the \ce{WO3} films have low mosaicity and present single crystal quality.
Performing reciprocal space maps (\cref{fig:WO3_RSM}) we find that the in-plane lattice of all films is coherently strained to the substrate one ($a_{\ce{WO3}}=a_{\ce{SrTiO3}}=\SI{3.905}{\angstrom}$).
By using Bragg's law, we extract in \cref{fig:caxis} the out of plane $c$-axis parameter from the $2\theta$ position of \ce{WO3} diffraction peaks.
At \po$=\SI{100}{\micro\bar}$ we obtain $c=\SI{3.70}{\angstrom}$.
Samples grown at lower \po present a larger $c$-axis parameter (\cref{fig:caxis}).
Because all films are epitaxially locked in-plane to the substrate lattice, this points to an increased \ce{WO3} unit cell volume, a trend compatible with a higher concentration of oxygen vacancies, as previously reported for oxygen deficient thin films grown by PLD \cite{
	mcintosh2006oxygen,
	choi2008oxygen,
	hauser2015correlation,
	lorenz2016correlation,
	enriquez2017oxygen
}.
For \po$<\SI{40}{\micro\bar}$ the c-axis becomes bigger than $a_{\ce{SrTiO3}}$, signalling a transition from tensile to compressive strain.
We note that, concomitantly with this crossover, the finite size oscillations disappear, suggesting that films grown at \po$<\SI{40}{\micro\bar}$ have a lower crystal quality.

By simulating the finite size fringes of the XRD data with a kinematic scattering model, we extract in \cref{fig:Nuc} the number of unit cells $N_{\ce{WO3}}$ forming the thin films. 
We find a constant $N_{\ce{WO3}}=\SI{28}{\unitcell}$ for \po$\leq \SI{60}{\micro\bar}$, and lower values at higher oxygen pressures.
Considering that all films were deposited with the same total number of laser pulses, we associate the decreased $N_{\ce{WO3}}$ with enhanced scattering of the plasma plume at higher pressures, which reduces the amount of material deposited on the substrate.
For \po$\leq \SI{20}{\micro\bar}$, the absence of finite size oscillations makes impossible to determine the number of unit cells by $\theta$--$2\theta$ measurements.
We thus perform X-ray reflectivity measurements (raw data in \cref{fig:XRR}) from which we extract the thin film thickness and interface roughness indicated by the crosses in \cref{fig:XRD_thickness,fig:XRD_roughness}.
The total film thickness shows an increasing trend with lower pressure, which is a combined effect of the $c$-axis expansion and decreased plume scattering.
In \cref{fig:XRD_thickness} we also evaluate the film thickness from the $\theta$--$2\theta$ measurements as $c\cdot N_{\ce{WO3}}$ (circles) and find good agreement with the reflectivity data.
Concerning the interface roughness in \cref{fig:XRD_roughness}, we find a minimum at \po$=\SI{40}{\micro\bar}$, consistent with the minimum value obtained from the topography data of \cref{fig:Roughness}.
These measurements show that \ce{WO3} $c$-axis parameter can be changed up to \SI{10}{\percent} by tuning the oxygen pressure during growth, while preserving the in-plane match with the substrate lattice and a flat surface.
It is then interesting to study how the film electronic properties are affected by \po, which we investigate by low-temperature transport measurements.

\subsection{Electrical transport}
\begin{figure}[tb]
\includegraphics[page=3,width=1\linewidth]{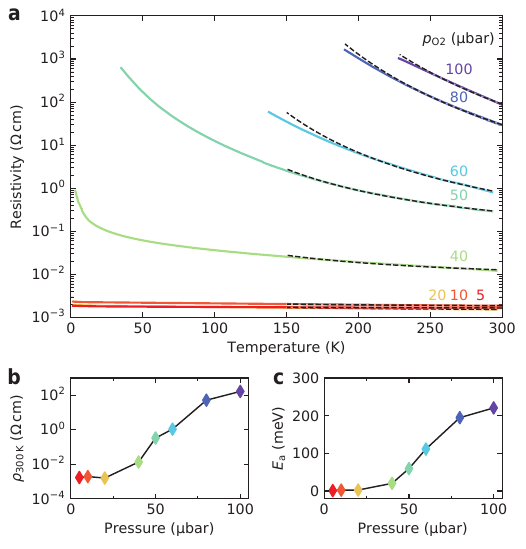}

\subfloat{\label{fig:WO3_RvsT}}
\subfloat{\label{fig:R300K}}
\subfloat{\label{fig:Egap}}

\caption{\textbf{Resistivity measurements.}
	\protect\subref{fig:WO3_RvsT} Four probe resistivity versus temperature curves.
	The dashed lines are best fits to the Arrhenius model for activated transport in \cref{eq:WO3_Arrhenius}.
	\protect\subref{fig:R300K} Room temperature resistivity and
	\protect\subref{fig:Egap} activation energy for transport extracted from the fits.
}

\label{fig:WO3_Transport}

\end{figure}

To investigate the effect of oxygen vacancies on \ce{WO3} electronic properties, we measure resistivity versus temperature curves on samples grown in different oxygen pressure conditions.
The measurements are performed in a van der Pauw configuration, and electrical contact to \ce{WO3} is obtained with ultrasonic bonded Al wires.
All the samples show a semiconducting behaviour (\cref{fig:WO3_RvsT}), with the resistivity monotonously increasing upon lowering temperature.
The resistivity $\rho$ is calculated using the film thickness measured by X-ray reflectivity.
Lower \po results in flatter curves with lower room-temperature resistivity, as highlighted in \cref{fig:R300K} where we report $\rho_\mathrm{300K}$ for the different samples.
The data shows a sigmoid trend, with a variation of more than 5 orders of magnitude in the studied range.
Such trend is comparable to the one observed for the $c$-axis parameter in \cref{fig:caxis}, suggesting the existence of a direct correlation between the lattice expansion and the electronic doping.
We describe the transport data using an Arrhenius-type behaviour
\begin{equation}
\label{eq:WO3_Arrhenius}
	\sigma (T) = \frac{1}{\rho(T)} \propto \exp{\left(-\frac{E_\mathrm{a}}{k_\mathrm{B}T}\right)},
\end{equation}
where $E_\mathrm{a}$ is the activation energy for charge transport.
The experimental curves are fit from room temperature down to \SI{150}{\kelvin} (dashed lines in \cref{fig:WO3_RvsT}), showing good agreement (fits on a larger temperature range are discussed in \cref{fig:A_VRH}).
We extract $E_\mathrm{a}\sim\SI{220}{\milli\electronvolt}$ at the highest \po, which is about one order of magnitude smaller than the optical band of \SI{3}{\electronvolt} found in bulk \ce{WO3} \cite{
	bullett1983bulk,
	sahle1983electrical
}.
This suggests that the observed activated transport arises from localized states lying inside \ce{WO3} band gap, which formed in a certain amount even at the highest \po value.
At lower \po, the concentration of oxygen vacancies in the film increases determining a vanishing $E_\mathrm{a}$.
Even though the \ce{WO3} films are on the verge of an insulator-to-metal transition, we measure semiconducting behaviour even at the highest vacancies concentration.
This is different from what observed with other doping mechanism, where a metallic state was achieved at high doping levels \cite{
	sahle1983electrical,
	haldolaarachchige2014superconducting,
	altendorf2016facet,
	yoshimatsu2016insulator
}.
Decreased crystal quality and disorder during low-pressure PLD growth are the most probable causes for the persistent semiconducting state measured in our most doped \ce{WO3} thin films.

To get indication on the density of the charge carriers, we perform low temperature magnetotransport measurements.
Because most of the samples are highly resistive, reliable Hall measurements could be obtained only for samples grown at \po$\leq \SI{20}{\micro\bar}$.
In this high-doping conditions, all the samples show similar magnetotransport data (\cref{fig:WO3_MT}), from which we extract a carrier density $n_\mathrm{3D}=\SI{4e21}{\per\cubic\centi\metre}$ and a mobility $\mu=\SI{0.6}{\squared\centi\metre\per\volt\per\second}$ at \SI{1.5}{\kelvin}, in agreement with previous reports of oxygen-deficient \ce{WO3} materials \cite{
	meng2015electrolyte,
	altendorf2016facet%
}.
The measured electron density corresponds to about $0.25$ electrons per unit cell, which would be equivalent to an oxygen vacancy concentration of about \SI{4}{\percent}, if these are the only source of electron donors.

\section{Ab initio calculations}
\begin{figure*}[tb]
\includegraphics[page=4,width=1\linewidth]{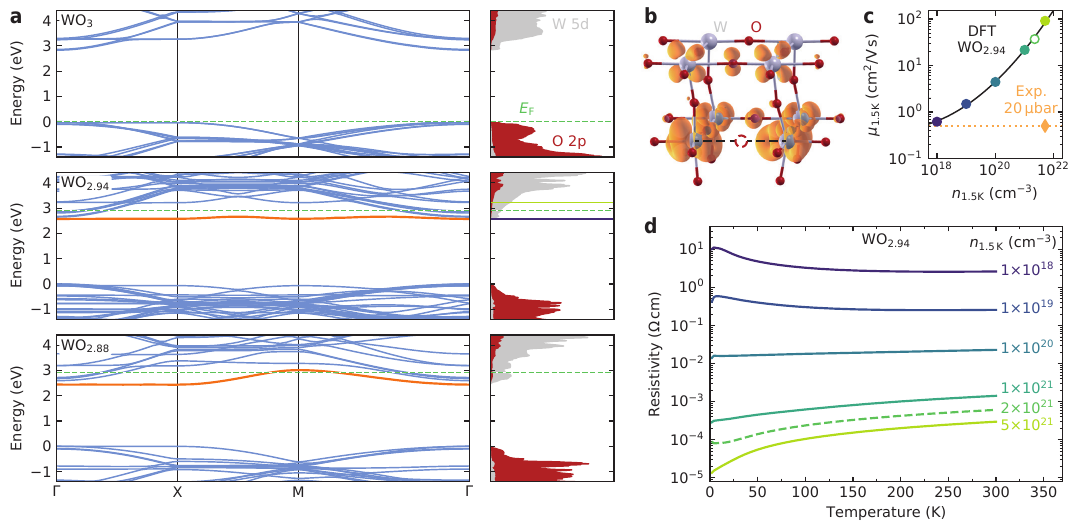}

\subfloat{\label{fig:DFT_Bands}}
\subfloat{\label{fig:DFT_Density}}
\subfloat{\label{fig:DFT_n_mu}}
\subfloat{\label{fig:DFT_RvsT}}

\caption{\textbf{DFT calculations of heteroepitaxial \ce{WO3} films with oxygen vacancies.}
	\protect\subref{fig:DFT_Bands} Left panels: in-plane band structure for \ce{WO3} with \SI{0}{\percent}, \SI{2}{\percent} and \SI{4}{\percent} of oxygen vacancies.
	The calculations are performed using 2x2x4 supercells, with an oxygen vacancy in the \ce{WO2} planes.
	The band of the oxygen defect states is indicated in orange, and the Fermi level is represented by the green dashed line.
	Right panels: density of states projected on the oxygen 2p (red) and tungsten 5d (grey) orbitals.
	For \ce{WO_{2.94}} the horizontal solid lines show the Fermi level corresponding to carrier densities $n_\mathrm{1.5K}=\SI{5e21}{\per\centi\metre\cubed}$ (green) and \SI{1e18}{\per\centi\metre\cubed} (violet).
	\protect\subref{fig:DFT_Density} Computed 2x1x1 supercell of \ce{WO_{2.94}} showing the real space charge density isosurface of the defect band (orange).
	The oxygen vacant site at the bottom is indicated by the dashed red circle.
	\protect\subref{fig:DFT_n_mu} Mobility as a function of carrier density (circles) at \SI{1.5}{\kelvin}, from the DFT calculations.
	The density at the Fermi level is indicated by the white circle.
	The diamond represents the experimental data from Hall effect measurements for the sample grown at \po$=\SI{20}{\micro\bar}$.
	\protect\subref{fig:DFT_RvsT} Resistivity versus temperature curves for \ce{WO_{2.94}} from the DFT calculations.
}

\label{fig:DFT}

\end{figure*}

\begin{table}[b]
\centering
\caption{\textbf{Ab initio results.}
	Properties from bulk $\delta$-\ce{WO3} and for films strained in-plane to the \ce{SrTiO3} substrate lattice with different concentrations of oxygen vacancies in the \ce{WO2} planes.
	The effective mass $m^*$ is calculated along the planar \ce{\Gamma-M} direction.
}
\label{tab:DFT}
\setlength{\tabcolsep}{6pt}     
\renewcommand{\arraystretch}{1.2} 
\begin{tabular}{r|cccc}
	& Bulk $\delta$-\ce{WO3} & \ce{WO3} & \ce{WO_{2.94}} & \ce{WO_{2.88}} \\
	\hline
	\ce{O} vac. & \SI{0}{\percent} & \SI{0}{\percent} & \SI{2}{\percent} & \SI{4}{\percent} \\
	$c$ (\SI{}{\angstrom}) & 3.75 & 3.685 & 3.72 & 3.73 \\
	$E_\mathrm{g}$ (\SI{}{\electronvolt}) & 2.71 & 2.83 & - & - \\
	$m^*$ ($m_\mathrm{e}$) & 0.56 & 0.85 & 2.41 & 1.16 \\
\end{tabular}

\end{table}
To investigate the origin of the observed structural and transport behaviour of \ce{WO3} films, we performed \textit{ab initio} calculations by variational pseudo-self-interaction-corrected (VPSIC) density functional theory (DFT) \cite{
filippetti2011variational,
archer2011exchange,
puggioni2012ordering,
delugas2013large
}.
To validate our results, we first calculated several \ce{WO3} bulk phases (further details in \cref{fig:WO3_structure} and the Supplemental Material).
The growth on \ce{SrTiO3} substrates was simulated using pseudo-cubic supercells, with in-plane lattice constant fixed at $a_{\ce{SrTiO3}}=\SI{3.905}{\angstrom}$ and the orthogonal axis left free to relax.
The calculations show that the stress associated with the planar expansion is partially relieved with a shrinkage of the out-of-plane lattice parameter to $c=\SI{3.685}{\angstrom}$ (\cref{tab:DFT}).
This value is comparable with the experimental data of \ce{WO3} films grown at the highest oxygen pressure reported in \cref{fig:caxis}, which are thus identified as stoichiometric.

Starting from the strained \ce{WO3} structure, we introduce vacancies by removing oxygen atoms in the \ce{WO2} crystal planes.
We study oxygen vacancy concentrations of \SI{2}{\percent} (\ce{WO_{2.94}}) and \SI{4}{\percent} (\ce{WO_{2.88}}).
After a further structural relaxation, our calculations find a lattice increase to $c=\SI{3.72}{\angstrom}$ and $c=\SI{3.73}{\angstrom}$ for the two vacancy concentrations.
This expansion is in qualitative agreement with our experimental results of \cref{fig:WO3_XRD}, even though the calculated values are considerably lower than what measured for films grown at low pressure.
Previous literature reports show that the lattice expansion due to oxygen vacancies is maximum when the charge is fully localised \cite{
marrocchelli2012charge,
marrocchelli2013chemical
}.
As discussed in the following, the structures described by our DFT calculations are in the metallic regime, and this may partially explain the quantitative discrepancy with the experiments.
Furthermore, temperature-related effects, which are not included in our calculations, could also affect the lattice expansion \cite{
glensk2014breakdown}.

In \cref{fig:DFT_Bands} we show the calculated in-plane band structure for strained \ce{WO3} films with different oxygen vacancy concentrations (extended plots in \cref{fig:Bands_supp}).
A single oxygen vacancy generates an excess of two electrons.
In the limit of an isolated point defect, this charge remains trapped by the \ce{W} atoms surrounding the vacancy, which change their ionic charge from the stoichiometric \ce{W^{6+}} to a lower value, which also gives rise to colour centres responsible for the photochromic effect \cite{
bange1999colouration,
deb2008opportunities
}.
Localisation may eventually be strengthened by the formation of small polarons \cite{
bange1999colouration
}.
According to our calculations, the excess charge reverses into the \ce{t_{2g}} orbitals of the \ce{W} atoms, forming a shallow band right below the bottom of the conduction band (orange band in \cref{fig:DFT_Bands}).
Although this band is optically separated from the bulk-like \ce{t_{2g}} conduction bands running above it, the system for both the examined doping levels is metallic, as indicated by the position of the Fermi energy (green dashed lines in \cref{fig:DFT_Bands}).
The defect state can be visualised in real space through the isosurface plot of \cref{fig:DFT_Density}: the defect charge is mostly localised on the two W atoms nearest neighbours of the vacancy, with clear \ce{t_{2g}} symmetry.
This is also seen by the broader bandwidth of the vacancy state along the diagonal direction \ce{X-M}, characteristic of \ce{t_{2g}} orbitals, instead of the cubic edge direction \ce{\Gamma-X}.

In \cref{tab:DFT} we report the band gaps and effective masses along the planar \ce{\Gamma-M} direction.
The vacancy-free structure shows an insulating state with a direct gap $E_\mathrm{g}=\SI{2.83}{\electronvolt}$ at the \ce{\Gamma} point, substantially larger than the indirect gap in bulk $\delta$-\ce{WO3} of \SI{2.71}{\electronvolt}.
This is a consequence of the tensile strain present in the structure which stretches the planar bonds and reduces the \ce{t_{2g}} bandwidth, as evidenced by the increased effective mass.
Even though the \ce{WO_{2.94}} and \ce{WO_{2.88}} structures have a small optical gap at $\Gamma$ of \SI{75}{\milli\electronvolt} and \SI{150}{\milli\electronvolt}, respectively, they show metallic character in transport due to the broader density of states in the orthogonal direction (\cref{fig:Bands_supp}).
With more excess charge the gaps and effective masses are progressively reduced, due to electronic correlation and band filling effects, which cause a spread in the bandwidth with respect to the empty conduction bands of the undoped system.
For lower vacancy concentration, up to the limit of isolated point defect, instead, the carrier density is expected to be reduced, leading to the formation of an insulating state.

In order to theoretically estimate electron mobility and resistivity in a wider range of vacancy concentrations, we apply the Bloch-Boltzmann approach to mix the \textit{ab initio} band energies of the \ce{WO_{2.94}} structure with a model description of the electron--phonon scattering \cite{
	filippetti2012thermopower,
	delugas2013doping
}.
We tune the Fermi level in a rigid band approximation to simulate the variation in the excess carrier density produced by the vacancies.
Results of three-dimensional averages of mobility versus carrier density at \SI{1.5}{\kelvin} are reported in \cref{fig:DFT_n_mu}.
As a consequence of the increasing population of high-energy \ce{t_{2g}} bands of \ce{WO3}, the mobility increases with the carrier density and, for \SI{2}{\percent} vacancies, we find $n=\SI{2.2e21}{\per\centi\metre\cubed}$ and $\mu = \SI{30}{\centi\metre\squared\per\volt\per\second}$ (white circle in \cref{fig:DFT_n_mu}).
In our oxygen deficient films, we measure a Hall mobility $\mu=\SI{0.5}{\centi\metre\squared\per\volt\per\second}$ (orange diamond in \cref{fig:DFT_n_mu} and raw data in \cref{fig:WO3_MT}) which, in our calculations, would be achieved for a considerably lower carrier density $n=\SI{1e18}{\per\centi\metre\cubed}$.
By using the \textit{ab initio} transport coefficients, we calculate the corresponding resistivity versus temperature in \cref{fig:DFT_RvsT}.
At densities below $n=\SI{1e19}{\per\centi\metre\cubed}$ we observe an insulating transport regime, for which only the energy-flat, lowest portion of the defect state is populated, with vanishing band velocities.
In this regime, the electrical transport is thermally activated, similarly to what we measure experimentally.
Above $n=\SI{1e19}{\per\centi\metre\cubed}$, an insulator-to-metal transition is observed in our calculations, but it is not achieved experimentally.
This is a clear indication that disorder in our thin films plays a crucial role in preventing occurrence of an insulator-to-metal transition, consistently with the experimental observation of a low electron mobility.

Several aspects of transport calculations are in good agreement with the measurements.
In particular, the amplitude of resistivity versus temperature matches well the experimental data, and the shallow temperature dependence of the resistivity is experimentally verified for most of the doped samples.
Furthermore, the calculations show that an insulator-to-metal transition occurs for $\rho_\mathrm{300K}<\SI{1e-2}{\ohm\centi\metre}$, a value consistent with our experimental observations.
On the other hand, there is an offset between calculated and Hall charge densities corresponding to the same mobility value, the former being two orders of magnitude lower.
This discrepancy suggests that a large portion of excess charge in the samples may be scarcely mobile, possibly stuck in deep traps or in highly localized polarons.
The hypothesis of additional charge in the samples coming from in-gap states is coherent with the experimentally measured Arrhenius trend of the resistivity.

\section{Conclusions}
In summary, we demonstrated that structural, electronic and transport properties of \ce{WO3} thin films can be controlled by changing the oxygen pressure during PLD growth.
The out-of-plane lattice constant of our thin films increases up to \SI{10}{\percent} as a consequence of the introduction of oxygen defects, while a pseudocubic phase of single crystal quality is maintained.
Oxygen vacancies act as electron donors and dope the material towards a metallic state.
Our VPSIC-DFT calculations describe oxygen vacancies in strained \ce{WO3} films as a weakly localized, shallow donors of \ce{t_{2g}} orbital character, lying only a few tents of \SI{}{\milli\electronvolt} below the bulk-like \ce{t_{2g}} mobility edge.
According to our Bloch-Boltzmann rigid-band calculations, full localization is only reached for $n<\SI{1e19}{\per\centi\metre\cubed}$, corresponding to an extremely small \SI{0.01}{\percent} vacancy concentration.
While several aspects of the calculations are coherent with the measurements, it is very likely that additional effects, like disorder and defect clusterization, can enhance charge localization, thus shifting to higher values the vacancy concentration threshold which separate charge-localized and delocalized regime.
Our results represent a fundamental step towards the understanding and engineering of a material which is likely destined to become a rising star in the energy and nanoelectronic applications of the future.

\section{Acknowledgments}
This work was supported by The Netherlands Organisation for Scientific Research (NWO/OCW) as part of the Frontiers of Nanoscience program (NanoFront) and by the Dutch Foundation for Fundamental Research on Matter (FOM).
The research leading to these results has received funding from the European Research Council under the European Union's H2020 programme/ ERC GrantAgreement n. [677458].

\bibliography{Biblio_WO3,Biblio_DFT,Biblio_Nickelates,Biblio_OtherOxides}


	\onecolumngrid
	\appendix
	\newpage
	\noindent\rule{1\columnwidth}{1pt}
	\section{\huge \texttt{Supplementary Information}}
	\noindent\rule{1\columnwidth}{1pt}

\renewcommand\thefigure{S\arabic{figure}}    
\setcounter{figure}{0}
\renewcommand\thetable{S\arabic{table}}    
\setcounter{table}{0}
\renewcommand\theequation{S\arabic{equation}}    
\setcounter{equation}{0}

\begin{figure*}[h]
	\includegraphics[page=5,width=.44\linewidth]{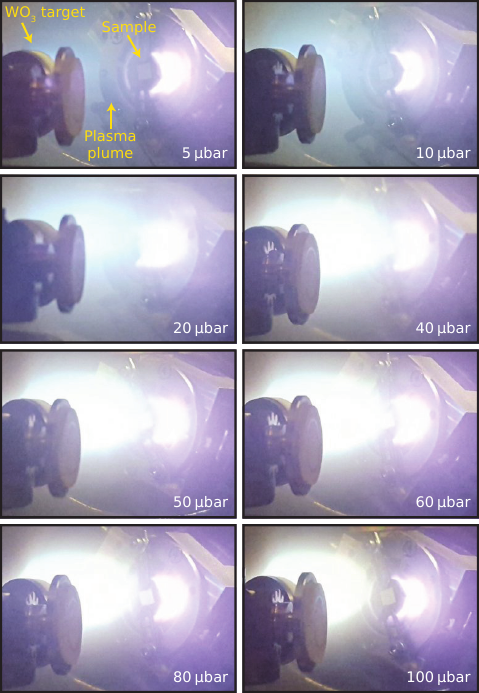}
	
	\caption{\textbf{Plasma plume of ablated \ce{WO3} at different oxygen pressures.}
		Photographs of the PLD vacuum chamber, showing the carousel of targets on the left (the \ce{WO3} target is facing backwards) and the sample holder on the right.
		The sample (square) is heated from the back with an \SI{800}{\nano\metre} continuous laser (saturating the camera).
		The pulsed laser radiation ($\lambda=\SI{248}{\nano\metre}$) enters the vacuum chamber from the bottom right and impinges on the rotating \ce{WO3} target, generating a plasma plume directed perpendicularly towards the sample.
		At the lowest \po the plume is barely visible, due to minimal scattering of the ablated species with the background gas.
		With increasing \po, the plume becomes more intense and confined.
		At \po$\ge \SI{80}{\micro\bar}$ the plume almost completely vanishes in intensity before reaching the sample, indicating the occurrence of many scattering events that ensure a high oxidation level of the material depositing on the substrate.
		This is consistent with our experimental observations in \cref{fig:WO3_XRD} of the main text, and also explains the decreased \ce{WO3} thickness which is deposited at the highest \po values, since more scattering events reduce the amount of material reaching the sample.
	}
	
	\label{fig:Plumes}
	
\end{figure*}

\begin{figure*}[h]
	\includegraphics[page=6,width=.7\linewidth]{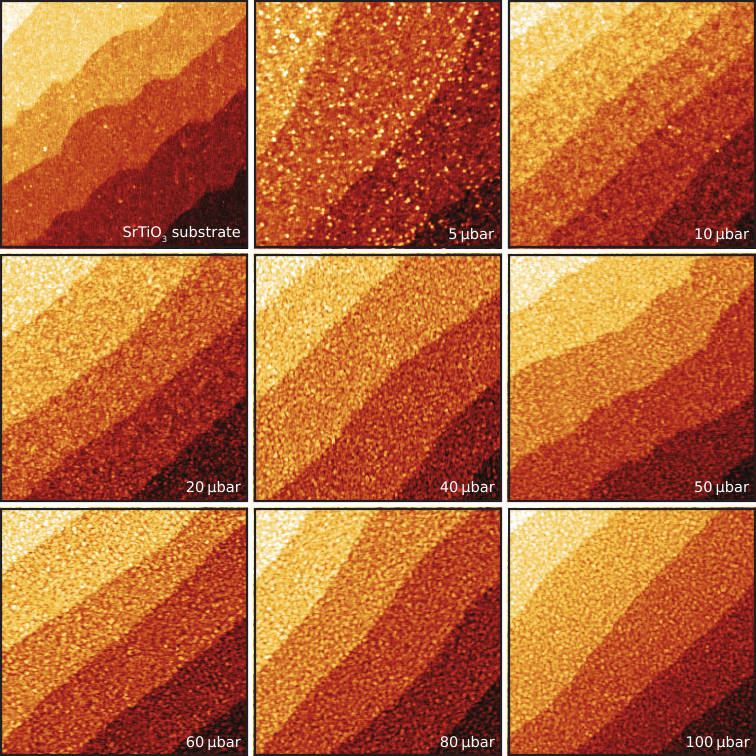}
	
	\caption{\textbf{Topography of the samples.}
		Atomic force microscopy images (\SI{2x2}{\micro\metre} in size) show the surface of the \ce{SrTiO3} substrate and of \ce{WO3} films grown in different oxygen pressures.
		A flat morphology with single unit cell steps and terraces is observed for all pressure conditions, with a minimum in surface roughness at \po$=\SI{50}{\micro\bar}$ (data in \cref{fig:Roughness} of the main text).
	}
	
	\label{fig:AFM_many}
	
\end{figure*}

\begin{figure*}[h]
	\includegraphics[page=7,width=.7\linewidth]{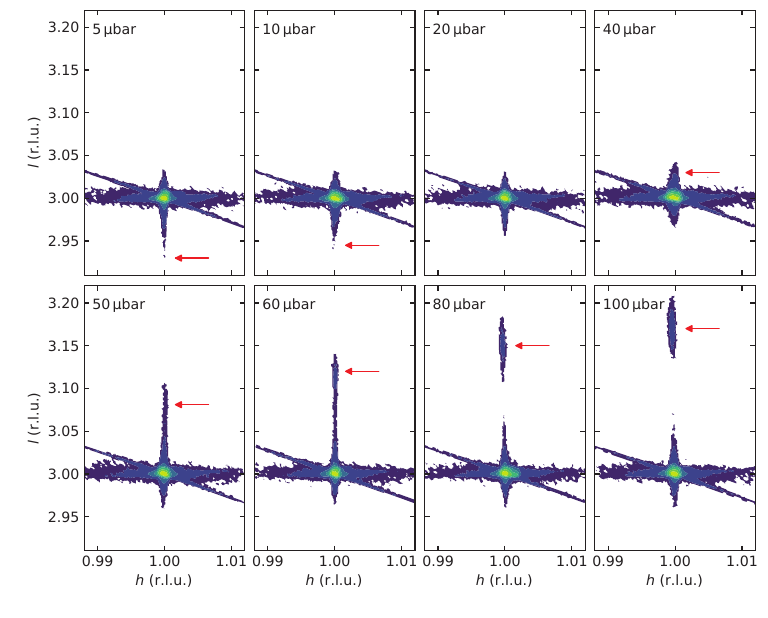}
	
	\caption{\textbf{Reciprocal space maps around \ce{SrTiO3} (103) diffraction peak.}
		The \ce{WO3} films determine a peak (indicated by the red arrows) which continuously changes from $l\sim2.93$ to $l\sim 3.17$ with increasing \po.
		For all growth pressure the \ce{WO3} films are coherently strained in-plane to the substrate lattice, as indicated by the identical alignment along the in-plane direction $h$.
	}
	
	\label{fig:WO3_RSM}
	
\end{figure*}

\begin{figure*}[h]
	\includegraphics[page=8,width=.7\linewidth]{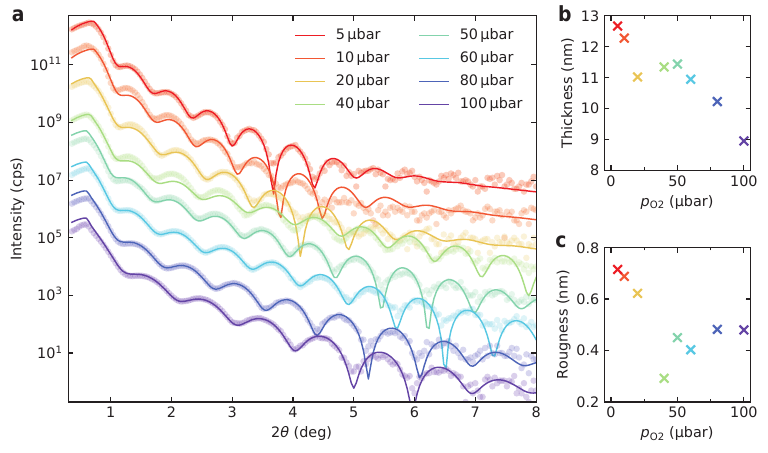}
	
	\subfloat{\label{fig:XRR_curves}}
	\subfloat{\label{fig:XRR_thickness}}
	\subfloat{\label{fig:XRR_roughness}}
	
	\caption{\textbf{X-ray reflectivity of \ce{WO3} thin films.}
		\protect\subref{fig:XRR_curves} Low-angle $\theta$--$2\theta$ reflectivity measurements (circles) fit with a conventional thin film reflectivity model (solid lines).
		\protect\subref{fig:XRR_thickness} Film thickness and
		\protect\subref{fig:XRR_roughness} interface roughness extracted from the fits.
	}
	
	\label{fig:XRR}
	
\end{figure*}

\begin{figure*}[h]
	\includegraphics[page=9,width=.7\linewidth]{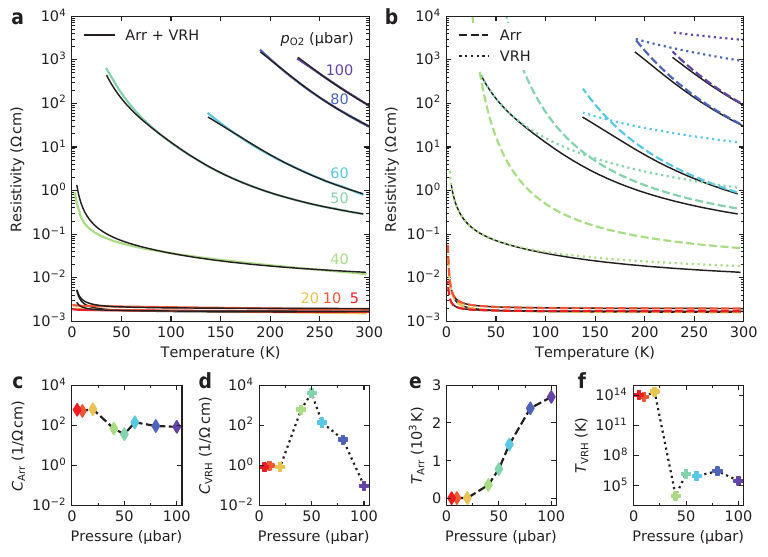}
	
	\subfloat{\label{fig:RvsT_A_VRH}}
	\subfloat{\label{fig:RvsT_A_VRH_separated}}
	\subfloat{\label{fig:RvsT_A_C}}	
	\subfloat{\label{fig:RvsT_VRH_C}}
	\subfloat{\label{fig:RvsT_A_T}}
	\subfloat{\label{fig:RvsT_VRH_T}}
	
	\caption{\textbf{Resistivity curves fit with Arrhenius and Mott variable range hopping.}
		\protect\subref{fig:RvsT_A_VRH} The experimental curves are fit with
		$\sigma(T) = 1/\rho(T) = C_\mathrm{Arr}\exp(-T_\mathrm{Arr}/T) + C_\mathrm{VRH}\exp(-(T_\mathrm{VRH}/T)^{1/4})$ (black lines) from room temperature down to $T=\SI{10}{\kelvin}$.
		This is a typical trend for electrical transport in complex oxide semiconductors [Catalan, G. et al., Phys. Rev. B 62, 7892-–7900 (2000)].
		Good agreement with the experimental data is observed at all temperatures.
		\protect\subref{fig:RvsT_A_VRH_separated} Separated contribution to the fit of the Arrhenius term (dashed lines) and variable range hopping term (dotted lines).
		The main resistivity contribution at $T\ge \SI{150}{\kelvin}$ is provided by the Arrhenius term, justifying the analysis performed in \cref{fig:WO3_Transport} of the main text.
		\protect\subref{fig:RvsT_A_C} Arrhenius and
		\protect\subref{fig:RvsT_VRH_C} variable range hopping scaling constants and
		\protect\subref{fig:RvsT_A_T},
		\protect\subref{fig:RvsT_VRH_T} corresponding temperature constants extracted from the fits.
		Note that the activation energy discussed in the main text can be calculated as $E_\mathrm{a} = T_\mathrm{Arr} k_\mathrm{B}$, where $k_\mathrm{B}$ is the Boltzmann constant.
	}
	
	\label{fig:A_VRH}
	
\end{figure*}

\begin{figure*}[h]
	\includegraphics[page=10,width=.7\linewidth]{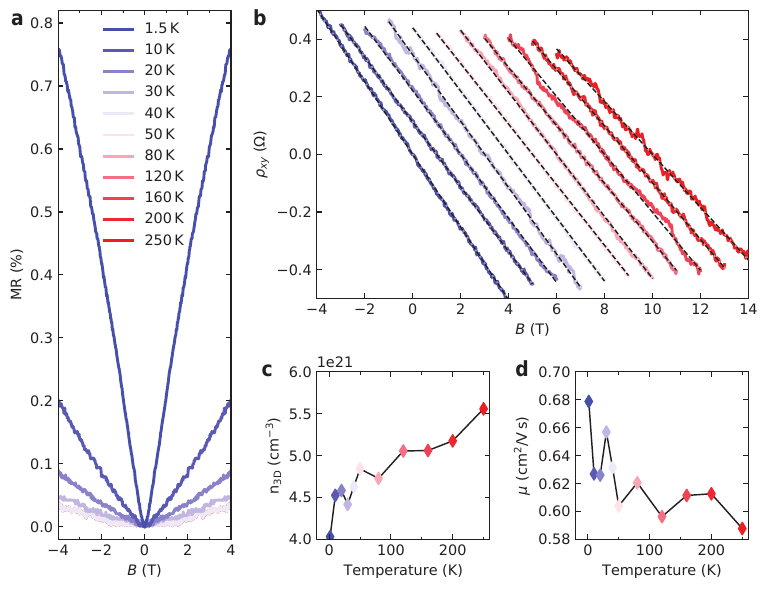}
	
	\subfloat{\label{fig:WO3_MT_MR}}
	\subfloat{\label{fig:WO3_MT_Hall}}
	\subfloat{\label{fig:WO3_MT_n}}	
	\subfloat{\label{fig:WO3_MT_mu}}
	
	\caption{\textbf{Magnetotransport measurements of \ce{WO3} grown at \SI{20}{\micro\bar}.}
		\protect\subref{fig:WO3_MT_MR} Magnetoresistance and
		\protect\subref{fig:WO3_MT_Hall} Hall effect curves measured at different temperatures (curves horizontally offset by \SI{1}{\tesla} for clarity).
		The magnetoresistance is defined as $\mathrm{MR}=\frac{\rho_\mathrm{xx}(B)-\rho_\mathrm{xx}(0)}{\rho_\mathrm{xx}(0)}$, where $\rho_{xx}$ is the longitudinal resistivity.
		A small positive magnetoresistance is observed, which becomes negligible above \SI{30}{\kelvin}, and it is reminiscent of low-temperature localisation effects.
		The Hall effect is negative and linear, indicating electron carriers, from which we extract
		\protect\subref{fig:WO3_MT_n} the three-dimensional carrier density and
		\protect\subref{fig:WO3_MT_mu} the mobility.
		At the lowest temperature we find $n_\mathrm{3D}= \SI{4e21}{\per\cubic\centi\metre}$, corresponding to 0.25 electrons per \ce{WO3} formula unit.
		Both $n_\mathrm{3D}$ and $\mu$ are almost constant in the studied temperature range.
	}
	
	\label{fig:WO3_MT}
	
\end{figure*}

\clearpage
\subsection{Additional ab initio calculations}
To validate our DFT results, we applied VPSIC to several \ce{WO3} bulk structures, for which an amount of experiments and earlier calculations is available.
As an example, for the triclinic $\delta$-\ce{WO3}, phase stable at room temperature, we obtain an indirect band gap of \SI{2.71}{\electronvolt} against a measured value of \SI{2.75}{\electronvolt}
[P.P. Gonz\'{a}lez-Borrero et al., APL 96, 061909 (2010)],
while for the low-temperature monoclinic $\varepsilon$-\ce{WO3} we have $E_\mathrm{g} = \SI{3.34}{\electronvolt}$, which compares well with $E_\mathrm{g} = \SI{3.27}{\electronvolt}$ obtained with the accurate, but computationally demanding, GW approach
[M.B. Johansson at al, J. Phys. Cond. Mat. 25, 205502 (2013)].
A comprehensive review with results for \ce{WO3} by a variety of advanced methods can be found in 
[F. Wang et al., J. Phys. Chem. C, 115, 8345--8353 (2011)].

\begin{figure*}[h]
	\includegraphics[page=11,width=.6\linewidth]{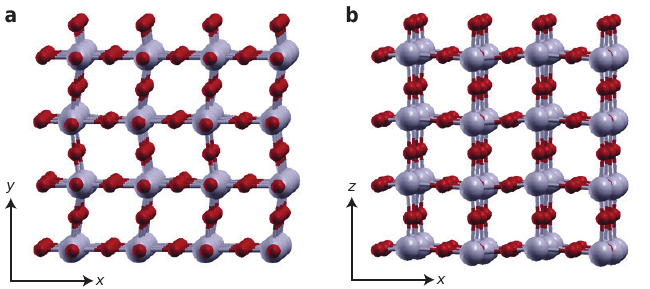}
	
	\subfloat{\label{fig:WO3_structure_xy}}
	\subfloat{\label{fig:WO3_structure_xz}}
	
	\caption{\textbf{Calculated structure for the \ce{WO3} film strained in the $(x,y)$ plane to match the \ce{SrTiO3} substrate lattice.}
		\protect\subref{fig:WO3_structure_xy} Film top view from the $(x,y)$ plane and
		\protect\subref{fig:WO3_structure_xz} side view in the growth direction from the $(x,z)$ plane.
		Grey and red balls are \ce{W} and \ce{O} atoms, respectively.
	}
	
	\label{fig:WO3_structure}
	
\end{figure*}

We simulated growth on the \ce{SrTiO3} substrate by using pseudocubic supercells with the in-plane lattice constant kept fixed at $a_{\ce{SrTiO3}}=\SI{3.905}{\angstrom}$, and the orthogonal axis free to relax.
Concerning atomic positions, we started the structural simulations from a hypothetical triclinic array with no symmetry imposed, and left all the atoms to relax towards the energy minimum.
Several supercell symmetries were considered, corresponding to different oxygen vacancy concentrations and configurations.
In \cref{fig:WO3_structure} we show the atomic structure obtained for the pristine strained film. 
Since the average lattice parameter in the $\delta$-\ce{WO3} phase is $a=\SI{3.75}{\angstrom}$
[P.M. Woodward et al., J. Phys. Chem. Solids 56, 1305--1315 (1905)],
we can estimate that the match with the substrate induces a linear tensile strain of about \SI{4}{\percent} in the film plane.
To partially relieve the stress associated with the planar expansion, the film shrunk longitudinally, with $c=\SI{3.685}{\angstrom}$.
Overall, the strained film displays a \SI{6}{\percent} volume expansion with respect to the $\delta$-\ce{WO3} bulk phase.
This value is compatible with the experimental lattice parameter of our most stoichiometric \ce{WO3} films $c=\SI{3.70}{\angstrom}$, within the \SI{1}{\percent} underestimation usually observed in DFT calculations
[Van de Walle, A. et al., Phys. Rev. B, 59(23), p.14992 (1999)],
[Haas, P. et al., Phys. Rev. B 79(8), p.085104 (2009)].

Another remarkable difference with respect to the bulk is in octahedral tiltings: in the planes (\cref{fig:WO3_structure_xy}) the \ce{W-O} bonds are stretched out and the octahedral tiltings largely suppressed, with \ce{W-O-W} angles of $\sim$\ang{175}, on average.
In the orthogonal direction (\cref{fig:WO3_structure_xz}) the \ce{W-O} bonds are more bulk-like, and the tiltings slightly larger (\ce{W-O-W} angles $\sim$\ang{168}).
The structure also displays octahedral distortions and shifts of W atoms from the octahedral centres, which are also present in the triclinic phase.
The overall tilting pattern is \ce{a^-b^-c^-}, like in $\delta$-\ce{WO3} bulk.

\begin{figure*}[t]
	\includegraphics[page=12,width=.7\linewidth]{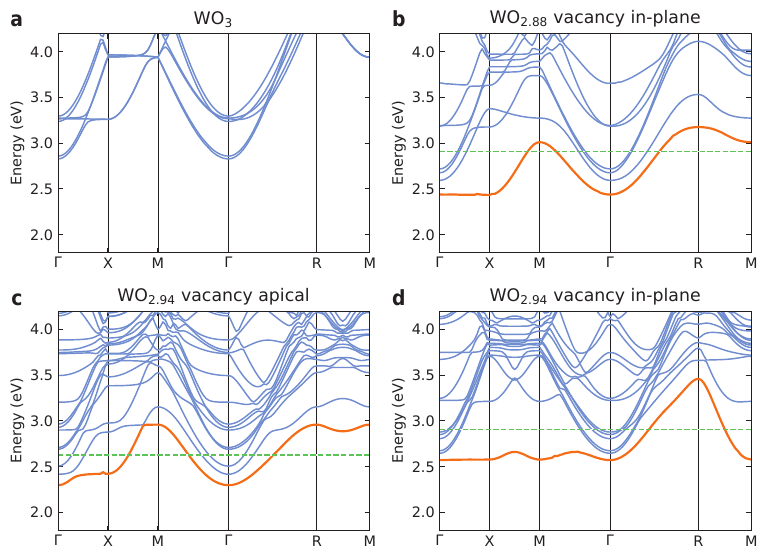}
			
	\subfloat{\label{fig:Bands_supp_WO3}}
	\subfloat{\label{fig:Bands_supp_WO288_inplane}}
	\subfloat{\label{fig:Bands_supp_WO294_apical}}
	\subfloat{\label{fig:Bands_supp_WO294_inplane}}
	
	\caption{\textbf{Band structure calculated for strained \ce{WO3} with and without oxygen vacancies.}
		\protect\subref{fig:Bands_supp_WO3} Pristine \ce{WO3} and with
		\protect\subref{fig:Bands_supp_WO288_inplane} \SI{4}{\percent} vacancies with oxygens missing in the \ce{WO2} planes,
		\protect\subref{fig:Bands_supp_WO294_apical} \SI{2}{\percent} vacancies in apical position and
		\protect\subref{fig:Bands_supp_WO294_inplane} \SI{2}{\percent} vacancies in-plane.
		The donor state resulting from the oxygen vacancies is drawn in orange.
		Zero energy is fixed at the valence band top (not shown) and the Fermi level of the neutral state is indicated by the green dashed lines.
		The $k$-points coordinates are $\Gamma=[0,0,0]$, $\mathrm{X}=[1/2,0,0]$, $\mathrm{M}=[1/2,1/2,0]$, $\mathrm{R}=[1/2,1/2,1/2]$ expressed in units of $[\pi/a, \pi/a, \pi/c]$, where $a$ and $c$ refer to the single \ce{WO3} unit cell.
	}
	
	\label{fig:Bands_supp}
	
\end{figure*}

In \cref{fig:Bands_supp} we show the calculated bands for the strained \ce{WO3} films.
The bands of the vacancy-free structure are visibly similar to that of the bulk $\delta$-\ce{WO3} phase, with two significant differences.
Firstly, $E_\mathrm{g} = \SI{2.83}{\electronvolt}$ is slightly larger than in bulk.
Secondly, our calculations show that in bulk $\delta$-\ce{WO3} the valence band at \ce{X} is \SI{3}{\milli\electronvolt} higher than at \ce{\Gamma}, resulting in an indirect \ce{X\bond{->}\Gamma} band gap.
For the strained film, instead, the valence band at \ce{\Gamma} is \SI{38}{\milli\electronvolt} higher than at \ce{X}, thus the gap is direct at the \ce{\Gamma} point.
We note that the band-gap nature for \ce{WO3} at room temperature is debated
[M.B. Johansson at al, J. Phys. Cond. Mat. 25, 205502 (2013)],
[F. Wang et al., J. Phys. Chem. C, 115, 8345--8353 (2011)],
although the indirect gap hypothesis seems the most credited by experimental reports
[H. Zheng et al., Adv. Funct. Mater. 21, 2175--96 (2011)],
[R. S. Vemuri et al., Appl. Mater. Interfaces, 4, 1371--1377 (2012)],
[D. G. Barton et al., J. Phys. Chem. B 103, 630--640 (1999)].
However, the flatness of the topmost valence band segment between X and \ce{\Gamma} suggests that minor structural differences in the sample could be sufficient to change the band gap nature.

Let us see more in detail these results: for \SI{4}{\percent} vacancies distributed in \ce{WO2} planes (\cref{fig:Bands_supp_WO288_inplane}) the defect state (in orange) is separated, at its energy-lowest \ce{\Gamma} point, by an optical gap of about \SI{150}{\milli\electronvolt} from the \ce{t_{2g}} bottom.
A large band dispersion of the defect state is visible both in plane (see along \ce{X-M}) and out of plane (along \ce{R-M}), indicating that vacancies separated by two lattice units are still affected by a fair amount of mutual interaction.
Overall the system is a semi-metal, that is optically insulating but metallic from the transport viewpoint.
Notice a significant detail: the broader bandwidth direction of the vacancy state is not along the cubic edge (\ce{\Gamma-X}) but along the diagonal \ce{X-M}).
This is a consequence of the \ce{t_{2g}} orbitals orientation, which do not point towards the nearest oxygen atoms, but diagonally.
In the attempt to further reduce the metallic character of the system, we tested two arrays with \SI{2}{\percent} vacancy concentration, one with vacancy in apical positions (\cref{fig:Bands_supp_WO294_apical}), another with vacancy in \ce{WO2} planes (\cref{fig:Bands_supp_WO294_inplane}).
In both cases the vacancy distance is doubled with respect to the \SI{4}{\percent} concentration, resulting in a substantial decrease of the bandwidth along the significant $k$-space direction (\ce{X-M} for in-plane, and \ce{R-M} for apical configuration).
For both cases the band bottom is still at \ce{\Gamma}, with optical gaps equal to \SI{120}{\milli\electronvolt} and \SI{75}{\milli\electronvolt} for apical and in-plane vacancies, respectively.
Even at this lower vacancy concentration, however, the fully localized regime is not reached.
The calculated effective masses give a measure of the localization degree of the defect state: for the in-plane vacancy state $m^* =2.41\,m_e$, while $1.01\,m_e$ in \ce{\Gamma-M} and \ce{\Gamma-R} directions, respectively.
These values can be compared with $m^* = 0.85\,m_e$ and $0.57\,m_e$ along the analogous directions for the lowest bulk-like \ce{t_{2g}} band.
The defect state is thus about 2--3 times less mobile than the states of the \ce{t_{2g}} conduction bottom.
For the apical vacancy state, we obtain $m^* =1.17\,m_e$ and $1.25\,m_e$ along \ce{\Gamma-M} and \ce{\Gamma-R}, which are not much larger than the  bulk-like \ce{t_{2g}} band values $m^* =0.86\,m_e$ and $0.74\,m_e$ in the corresponding $k$-space directions.

\end{document}